\documentclass[twocolumn,showpacs,preprintnumbers,amsmath,amssymb]{revtex4}

\usepackage{graphicx}
\usepackage{dcolumn}
\usepackage{bm}

\begin{document}

\title{Three-dimensional carrier-dynamics simulation of terahertz emission from photoconductive switches}
\author{E. Castro-Camus}
\email{e.castro-camus1@physics.ox.ac.uk}
\author{J. Lloyd-Hughes}
\author{M.B. Johnston}
\affiliation{University of Oxford, Department of Physics, Clarendon
Laboratory, Parks Road, Oxford OX1 3PU, United Kingdom}

\parbox{17.2cm}{\centering {\footnotesize \hfill PHYSICAL REVIEW B {\bf 71}, 195301 (2005) \hfill \tiny also appears in VIRTUAL JOURNAL OF ULTRAFAST SCIENCES (JUNE 2005)}}
\received{22 November 2004} \accepted{22 February 2005} \published{3
May 2005}

\begin{abstract}
A semi-classical Monte Carlo model for studying three-dimensional
carrier dynamics in photoconductive switches is presented. The model
was used to simulate the process of photo-excitation in GaAs-based
photoconductive antennas illuminated with pulses typical of
mode-locked Ti:Sapphire lasers. We analysed the power and frequency
bandwidth of THz radiation emitted from these devices as a function
of bias voltage, pump pulse duration and pump pulse location.  We
show that the mechanisms limiting the THz power emitted from
photoconductive switches fall into two regimes: when illuminated
with short duration ($<40\,\rm{fs}$) laser pulses the energy
distribution of the Gaussian pulses constrains the emitted power,
while for long ($>40\,\rm{fs}$) pulses, screening is the primary
power-limiting mechanism. A discussion of the dynamics of bias field
screening in the gap region is presented. The emitted terahertz
power was found to be enhanced when the exciting laser pulse was in
close proximity to the anode of the photoconductive emitter, in
agreement with experimental results. We show that this enhancement
arises from the electric field distribution within the emitter
combined with a difference in the mobilities of electrons and holes.
\end{abstract}

\maketitle

\section{Introduction} \label{s:introduction}

Over the last few years interest in using far-infrared radiation
as a non-contact probe of the properties of various materials such
as semiconductors \cite{HuberTBBAL01, LeitenstorferHTBBA02,
KaindlHCLC03}, superconductors
\cite{NussGGMOB91,GaoWLUPK95,FedericiGSDSD92} and biological
tissue \cite{HanCZ00, PickwellCFWP04} has grown considerably. This
has been as a result of improvements in sources \cite{ShenULBD03,
JohnstonWCDL02} and detectors \cite{ShenUBLDGBTE04} of terahertz
(THz) frequency radiation, and the development of novel
spectroscopic methods such as time domain spectroscopy
\cite{FergusonZ02, ShanNH02}. By recording the electric field of a
single cycle of terahertz radiation as function of time, the
technique of time domain spectroscopy is capable of extremely
sensitive measurements, especially in the far-infrared band. This
technique represents a particularly powerful tool for studying the
dynamics of weakly correlated systems in condensed matter physics,
including excitons and Cooper pairs, which have binding energies
observable in the THz region of the electromagnetic spectrum
\cite{KaindlHCLC03,FedericiGSDSD92}.

Terahertz time domain spectroscopy relies on the ability to
produce pulses of THz radiation so short that they contain only
one or even half a cycle. Single-cycle THz pulses can be generated
by shining ultrashort laser pulses on bulk semiconductors such as
InAs, GaAs and GaP \cite{HeldKM91,JohnstonWCDL02}. These pulses
contain a continuous and broad distribution of frequencies, which
can be exploited to perform spectroscopic studies over a large
portion of the spectrum.

In bulk semiconductors single-cycle radiation can be emitted via
three mechanisms: the formation of an electrical dipole due to the
difference of mobilities between electrons and holes (the
photo-Dember effect) \cite{JohnstonWCDL02}, optical rectification
\cite{AustonC85}, and the acceleration of charges under the
surface depletion field of a semiconductor \cite{JohnstonWCDL02}.
Photoconductive switches (PCSs) are an attractive alternative to
surface field semiconductors emitters for the production of
single-cycle THz radiation \cite{SmithAN88}. These devices have
been shown to produce broad bandwidth THz signals with high
electric field intensities \cite{ShenULBD03}. In a PCS two
parallel metallic contacts are deposited on the surface of a
semiconductor, an external biasing voltage is applied between
these contacts and an ultra-short laser pulse is used to generate
photo-carriers in the gap between the contacts. The acceleration
of these carriers by the external electric field gives rise to an
electromagnetic emission according to Maxwell's equations.

An understanding of the carrier dynamics in photo-conductive
switches is of major importance in order to improve the bandwidth
and power of these devices. Several theoretical and experimental
studies have been performed to understand different aspects of the
process of THz emission by PCS. These works include the analysis
of charge transport by the use of rate equations
\cite{RodriguezCT94,DekorsyPKK93} and
 Drude-Lorentz models \cite{PiaoTS00,JepsenJK96}. In addition, the effect of
dislocations on the emitted signal has been studied by the
inclusion of reduced mobilities in rate equations \cite{LiuQ03} as
well as experimentally measuring time domain THz traces and
fitting them to simple charge transport models
\cite{NemecPKKSW01}. Also, TDS experiments using double pulses to
excite the PCS \cite{SiebertLLR04} have been performed to obtain
information about the photo-generated carrier screening of the
bias voltage.

These models correctly predict most of the characteristics of the
observed THz signals, but have the disadvantage that parameters
such as mobility and carrier trapping time have to be adjusted
phenomenologically to reproduce the details of the experimental
measurements. While such models have the advantage that under
certain circumstances they can be analytically solved, this
possibility is limited by factors such as the geometry of the
domain in which the equations can be solved, and in some cases the
analysis has to be restricted to one or two dimensions
\cite{DekorsyPKK93,NemecPKKSW01,RodriguezT96}. A Monte Carlo
simulation for PCSs was recently published \cite{LiuQ03m} that
models the dynamics of a low density of electrons generated by
small photon fluences. However, it neglects the contribution of
holes to the current density (see Section \ref{s:theory}) as well
as the carrier-carrier scattering mechanism which has been found
to dominate the total carrier scattering rate
\cite{Lloyd-HughesCFJJ04}.

In this work a three-dimensional Monte Carlo model is presented
(Section~\ref{s:theory}) that provides a powerful tool for the
analysis of the influence of parameters on carrier dynamics and
the evolution of variables in time. The effect of the applied bias
voltage (Section~\ref{s:bias}) and pump pulse width
(Section~\ref{s:pulsewidth}) on the emitted signal is analyzed. A
discussion of electric field screening is also presented
(Section~\ref{s:screening}).

\section{ Monte Carlo Model } \label{s:theory}

In this paper we extend the Monte Carlo model of
Ref.~\cite{JohnstonWCDL02} to simulate emission from PCSs. The
model uses the current density vector $\mathbf{J}$ as source term
of Maxwell's equations. The electric field is calculated by the
use of the far-field approximation\begin{equation} \mathbf{E}_{\rm
THz}\propto \frac{{\rm d}\mathbf{J}}{{\rm d}t}. \label{e:farfield}
\end{equation}  In order to obtain the current
density vector, the simulation volume is divided into a
three-dimensional rectangular grid, and the displacement of each
carrier (electron or hole) is calculated over small time
intervals, which are shorter than the scattering times. In each
time interval and spatial box, the electric field is assumed to be
constant and uniform and then the displacement of each carrier is
calculated by assuming it follows the solution of
\begin{equation}
\mathbf{\ddot{r}}=\frac{q\mathbf{E}}{m^{*}},\label{e:newton}
\end{equation}
with initial conditions given by its previous position and velocity, where $%
\mathbf{\ddot{r}}$ is the acceleration of the carrier with charge
$q$ and effective mass $m^{*}$ in an electric field $\mathbf{E}$.

Although the simulation assumes parabolic band dispersion,
inclusion of both $\Gamma $ and L valleys creates a more realistic
carrier ensemble. A set of quantum mechanical scattering rates are
calculated. Random numbers are used to determine if each particle
is scattered in each time step, the scattering angle and the
energy loss. The scattering mechanisms included are: LO phonon
absorption and emission within both $\Gamma $ and L valleys, TO
phonon emission and absorption between $\Gamma $ and L valleys,
acoustic phonon, impurity and carrier-carrier scattering. Details
of the implementation of these scattering mechanisms are described
in Ref.\cite{JohnstonWCDL02}.

\begin{figure}
\begin{center}
\includegraphics[width=8.6cm]{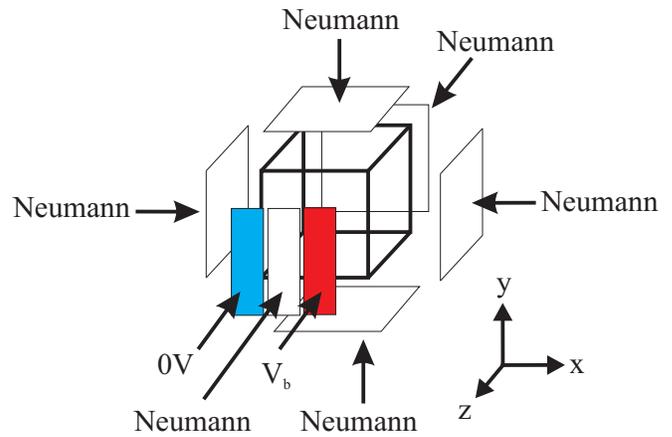}
\end{center}
\caption{ (Colour online) Diagram of the boundary conditions used. On the plane $z=0$ the boundary conditions are set as $0$ for $y<10\mathrm{\mu m}$, $V_{b}$ for $y>20\mathrm{%
\mu m}$ and Neumann boundary condition is assumed in the region $10\mathrm{%
\mu m}<y<20\mathrm{\mu m}$ as well as on all other boundaries.}
\label{Potential}
\end{figure}

In each time interval the charge density is calculated for each of
the boxes in the spatial grid. This is then used to find the
electric potential $\phi $ by numerically solving Poisson's
equation in three dimensions using a relaxation method with
Chebyshev acceleration.  The boundary conditions for solving
Poisson's equation are set as follows (Fig.~\ref{Potential}): the
potential is fixed to zero on one third of the top surface
($z=0$), corresponding to one of the contacts, on the second third
of the surface Neumann boundary conditions are assumed, and the
potential is fixed to $V_{\rm b}$ (the bias voltage) in the final
third. Neumann boundary conditions are also assumed on the other
boundaries. The electric-field vector $\mathbf{E}=-\nabla \phi $,
is determined using the central difference method. Initially the
spatial distribution of extrinsic carriers is calculated, and the
temperature distribution of these cold carriers is set according
to Maxwell-Boltzman statistics. The simulation is run until charge
distribution equilibrium is reached, prior to the photo-excitation
process.

Photo-excited carriers are assumed to be generated according to
the spatial and temporal intensity of a Gaussian laser pulse
absorbed in the sample. Random numbers are used to calculate the
initial positions of each photo-generated electron-hole pair. The
distribution in the $z$-direction is weighted exponentially as
defined by the absorption coefficient of the light in the
semiconductor. The initial energy of the electron-hole pairs is
the difference between the semiconductor band gap and the photon
energy, this energy is partitioned between electrons and holes
based on momentum conservation. Photons are assumed to have the
energy distribution of a Gaussian pulse (thus spectral width in
the simulation is calculated from the laser pulse duration
parameter). The directions of the pairs of momentum vectors are
initially randomized with equal probability across $4\pi $
steradians.

The \emph{net} current density vector is then calculated as

\begin{equation}
\mathbf{J}=\sum_{i}q_{i}\mathbf{\dot{r}}_{i}, \label{e:Jtotal}
\end{equation}
its derivative is calculated and with it the radiated electric
field using the far field approximation
(Equation~\ref{e:farfield}).

The parameters used in the simulation are taken from
Ref.~\cite{JohnstonWCDL02}. These parameters are typical of
insulating GaAs (an n-type sample with very low donor density of
$n=2\times10^{15}\,\rm{cm^{-3}}$ was chosen).  A 4\,pJ laser pulse
with $1.5\,\rm{\mu m}$ standard deviation beam-waist was simulated,
where the centre wavelength of the laser pulse was chosen to be
800nm (1.55\,eV). The domain of the simulation was $30\mathrm{\mu
m}\times 30\mathrm{\mu m}\times 5\mathrm{\mu m}\times 2\mathrm{ps}$,
divided in to a space-grid of $64\times 64\times 32$ and time
intervals of $2\,\mathrm{fs}$. The distance between the contacts of
the simulated photoconductive switch was $10\,\mathrm{\mu m}$. One
million pairs of pseudo-particles were used to represent extrinsic
and intrinsic electrons and holes.

\section{ Bias voltage effects} \label{s:bias}

\begin{figure}
\begin{center}
\mbox{\includegraphics[width=8.6cm]{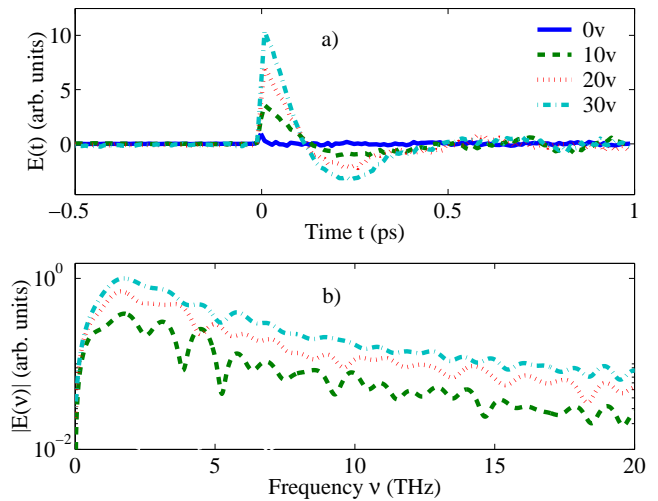}}
\end{center}
\caption{ (Colour online) Graphs of the THz waveform (upper panel)
for 10fs pump pulses at different voltages, and their Fourier
transforms (lower panel). The width of the pulse decreases with
voltage and this effect gives as a result an increase of the
bandwidth.} \label{10and100fs}
\end{figure}

\begin{figure}
\begin{center}
\mbox{\includegraphics[width=8.6cm]{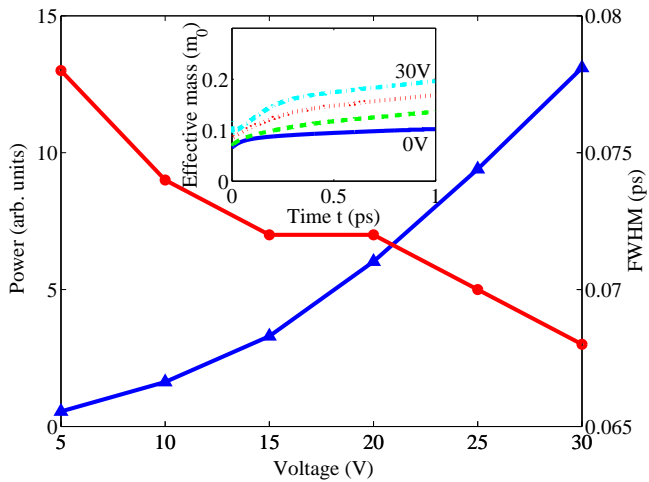}}
\end{center}
\caption{ (Colour online) Graph of full widths at half maximum
(circles) for $10\,\rm{fs}$ pump pulses as a function of the bias
voltage showing the narrowing of the THz transient with voltage. The
power of the signal emitted (triangles) is also shown as function of
applied voltage. The inset shows the evolution of the effective mass
of electrons as a function of time for 0, 10, 20 and 30V of bias
voltage.}\label{10and100wodths}
\end{figure}

Using the model described in the previous section we performed a
study of the effect of applied bias voltage on emitted THz power
from GaAs PCSs. Since mode-locked Ti:Sapphire
lasers are most commonly used in experiments we chose pulse
parameters typical of this type of laser. The simulation results
presented in Fig.~\ref{10and100fs}a show the effect of altering
the bias across a PCS. In this set of simulations the laser pulse
duration was set to 10\,fs, which is a pulse duration easily
achieved using chirped mirror Ti:Sapphire oscillators. The THz
pulse shortens at higher bias voltages as seen in Fig.~\ref{10and100wodths}, in which the FWHM of the pulses at different
voltages are plotted. This is because at higher applied voltages a
greater number of electrons are transferred into the L-valley,
resulting in a higher TO-phonon emission and absorption rate. As a
consequence the average effective mass of the electron population
increases (see inset in Fig.~\ref{10and100wodths}), reducing the
net electron mobility, thereby shortening the pulse. This effect
has been observed in the experiments of Liu \emph{et al.}
\cite{LiuTP03}. In Fig.~\ref{10and100fs}b the Fourier transforms
of Fig.~\ref{10and100fs}a are shown. These spectra broaden as
the voltage increases, produced by the shortening of the THz
transient.

The power emitted increases with the bias voltage, as seen in
Fig.~\ref{10and100wodths}, and in agreement with the
experimental measurements of \cite{LiuTP03}. This power
enhancement is expected in a simple \emph{macroscopic} picture.
Taking the current vector $\mathbf{J}=\sigma \, \mathbf{E}_b$,
with $\sigma$ being the switch's conductivity and $\mathbf{E}_b$
the external bias field, assuming that $\mathbf{E}_b$ remain
constant we get $\mathbf{E}_{\rm{THz}}\propto \dot{\sigma} \,
\mathbf{E}_b$. If the power is given by \begin{equation} P \propto
\int_{-\infty}^{+\infty}|\mathbf{E}_{\rm{THz}}|^2\,\rm{d} t,
\label{e:power}
\end{equation}
and since $|\mathbf{E}_b| \propto \rm{V}_b$ we end up with $P
\propto \rm{V}_b\,^2$, which is consistent with the quadratic-like
behaviour of power in Fig.~\ref{10and100wodths}. Note that the
assumption of $\mathbf{E}_b$ being constant, while a good first
approximation, it is not strictly true as shown by the further
analysis of the full \emph{microscopic} simulation (Section
\ref{s:screening}).

\section{ Pulse width effects } \label{s:pulsewidth}

The excitation pulse duration is one of the main variables
affecting the temporal width (and hence the frequency bandwidth)
of an emitted THz transient from a photoconductive switch. This is
primarily because the pulse duration determines the time interval
over which photo-carriers are generated, and therefore the current
rise time in a PCS. Here we present a set of simulations where the
excitation pulse duration is varied over a range achievable using
Ti:Sapphire lasers (6 to 120\,fs).

\begin{figure}
\begin{center}
\mbox{\includegraphics[width=8.6cm]{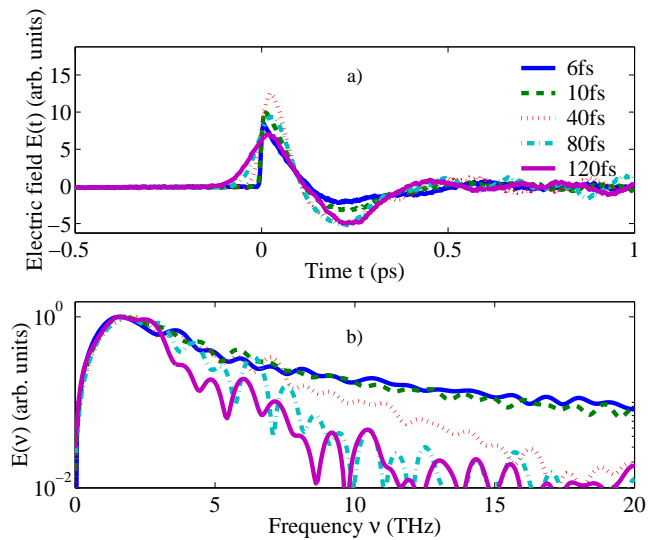}}
\end{center}
\caption{ (Colour online) Graphs of THz waveforms (upper panel) for
different widths of the pump pulse durations. The THz pulse sharpens
as the pump pulse width becomes smaller. The spectra of the previous
signals is shown in the lower panel, there is a significant effect
on the high frequency components.} \label{Pulses}
\end{figure}

\begin{figure}
\begin{center}
\mbox{
\includegraphics[width=8.6cm]{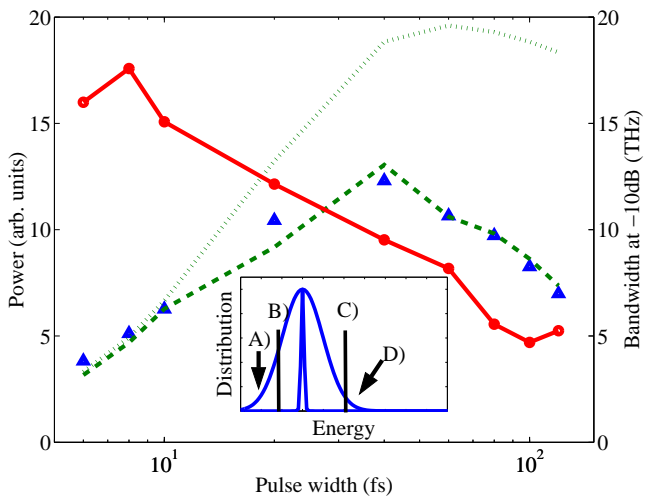}}
\end{center}
\caption{ (Colour online) Graph of the spectral width at -10dB as a
function of pump pulse width (circles) at a bias voltage of
$30\,\rm{V}$. The width of the distribution increases as the pump
pulse decreases caused by carriers being photo-generated in a
shorter period of time. The power of the signal emitted is also
plotted (triangles) as function of the pump pulse duration, the
dotted curve shows the power calculated taking into account the
number of photo-generated carriers and their effective mass, the
dashed line includes additionally the effective bias electric field.
The inset is a schematic with two curves (not drawn to scale) of the
energy distributions of a short and a long pulse: A) are the photons
with not enough energy to transfer carriers into the conduction
band, B) is the band-gap energy, C) is the band-gap energy plus the
L-valley offset and D) are the photons with enough energy to inject
carriers directly into the L valley} \label{PulseWidth}
\end{figure}

In Fig.~\ref{Pulses} the electric fields for various laser pulse
durations are shown for a bias voltage of $30\,\mathrm{V}$. It can
be seen that the THz transient becomes narrower as the pump pulse
gets shorter (Fig.~\ref{Pulses}a). In Fig.~\ref{Pulses}b the
spectra of the time-domain pulses are shown. While the
distributions for long pump pulses ($\sim120\,\rm{fs}$) show a
sudden drop in spectral components over $4\,\rm{THz}$,
distributions for short pulses ($\sim10\,\rm{fs}$) contain
components over $8\,\rm{THz}$. The distribution width at -10dB of
the spectra is used as a measure of the bandwidth of signal
generated in the PCS in Fig.~\ref{PulseWidth}. This plot shows
how the bandwidth of the signal increases as the pump pulse
shortens.

\emph{A priori} the power is not expected to depend significantly
on the exciting pulse width, as the number of photons per pulse
and the applied bias voltage are identical in the simulation. In
Fig.~\ref{PulseWidth} the power is plotted (triangles) as
function of the pump pulse width. Surprisingly a strong dependence
can be seen, with the power dropping for both over and under
$40\,\rm{fs}$ pulses. The drop for under $40\,\rm{fs}$ pulses can
be explained \emph{macroscopically} by calculating the current
density vector, given by
\begin{equation}
\mathbf{J}= (ne\mu_e+pe\mu_h)\mathbf{E}_b. \label{e:J}
\end{equation}
Taking the number densities $n$ and $p$ to be proportional to the
total number of carriers $N_e$ and $N_h$, extracted from the
simulation output, and calculating the mobilities $\mu_e$ and
$\mu_h$ from the average effective mass, which is
\begin{equation}
m_{\rm{eff}}=\frac{N_{\Gamma }m_{\Gamma }+N_{L}m_{L}}{N_{\Gamma
}+N_{L}} \label{e:meff}
\end{equation}
for electrons and $m_h$ for holes, the current vector can be
estimated, and with it the emitted power, $N_{\Gamma }$ and
$N_{L}$ were again extracted from the simulation. In Fig.
\ref{PulseWidth} the dotted line shows the expected power using
this method; the drop in power in the short pulse region
approximately reproduces that of the simulation. It can therefore
be attributed to the broadening of the energy distribution of the
pump pulse as its duration gets shorter, having as a consequence
two main effects. Firstly, the lower energy tail of the
distribution falls below the band-gap energy (see inset in Fig.~\ref{PulseWidth}), resulting in a smaller density of
photo-carriers generated. Secondly the high energy tail of the
pump pulse distribution contributes to the power reduction, since
a larger number of electrons are directly injected into the L
valleys, resulting in a larger electronic effective mass and thus
a lower mobility.

In order to explain the drop in power for long pulses, it is
necessary to take into account the actual electric field acting on
the photo-generated carriers. This can be done by replacing the
electric field $\mathbf{E}_b$ in Equation \ref{e:J} by an
effective electric field obtained by averaging the simulated
electric field over the size of the laser spot ($\sim
\pm0.8\,\rm{\mu m}$ from the center of the PCS's gap), and over
$\sim 100\,\rm{fs}$ (the screening time, see Section
\ref{s:screening}). The dashed line in Fig.~\ref{PulseWidth}
shows the power calculated using this effective electric field.
The good agreement between the simulations and this curve leads to
the conclusion that the drop in power for long pulses can be
attributed to the screening of the bias field.

\section{Screening} \label{s:screening}
\begin{figure}
\begin{center}
\mbox{\includegraphics[width=8.6cm]{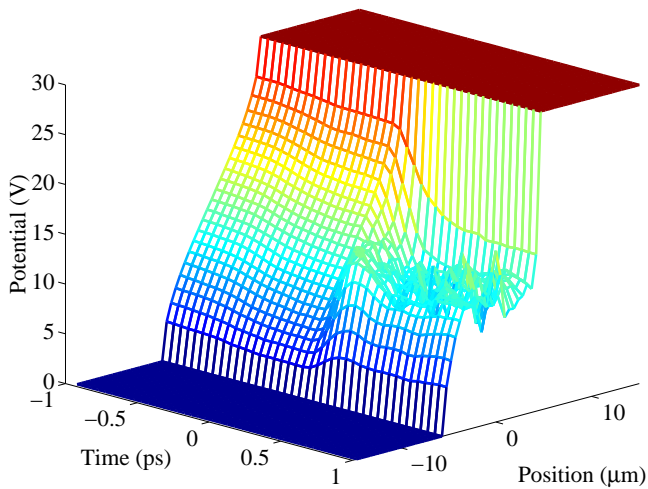}}
\end{center}
\caption{ (Colour online) Graph of electric potential as function of
$x$ and $t$. The surface show the screening of the electric field in
the gap region after the arrival of the pump pulse (at $t=0$) as the
separation of electrons and holes forms a dipole.} \label{Screening}
\end{figure}

The power and bandwidth of radiation emitted by a PCS is influenced
strongly by electric field screening effects. Therefore it is not
surprising that space-charge screening has been a major focus of
models describing PCS emission.\cite{ShanH04}

Even prior to photoexcitation of a PCS, the steady-state distribution
of extrinsic carriers around the electrodes must be considered in
order to determine the complete electric vector-field within the
active region of the PCS. In the present study this steady-state
space-charge field is automatically calculated (before the optical
excitation process is simulated) by iterating the simulation over
enough time-steps until an equilibrium electric potential is
reached. The form of the screened electric field between the
electrodes for an n-type GaAs PCS may be inferred from the gradient of
the $t=-1$\,ps trace in Fig.~\ref{Screening}. It can be seen that the extrinsic electron distribution is such that the
anode and cathode are screened to similar extents.

Here we focus on the r\^{o}le of space-charge screening during and after
the arrival of the pump pulse. On photo-excitation electron--hole pairs
are generated within the high electric field gap region of a
PCS. Charge separation produces a dipole, resulting in the emission of
THz radiation.  However, the resulting space-charge electric field
opposes the applied field, thereby reducing the force experienced by
carriers. It is expected that this effect would be
particularly significant for carriers photo-injected in the center of
the PCS gap at later times.

In Fig.~\ref{Screening} the simulated potential of the PCS is
plotted (for $z=0$) as a function of $x$-position and time. Before
the arrival of the pulse (at 0\,ps) the gradient of the potential
shows a continuous and positive $x$-component in the region of the
gap between the electrodes. After the injection of photo-carriers
the potential considerably flattens in the central part of the gap region,
corresponding to screening of the electric field.

It can be seen from Fig.~\ref{Screening} that an abrupt change in
the potential is present near the edges of the electrodes, which
increases as a result of the photo-carrier-induced
screening. Therefore, the gradient of the potential will have a large
$x$-component, implying a high electric field zone. If the
photo-carriers are generated near the electrodes an enhancement of the
emitted signal may be expected.

Experimental studies have shown that the power emitted by a PCS may be
enhanced by illuminating it asymmetrically. The enhancement is
greatest if the region illuminated with the pump-pulse is close to the
edge of the anode contact\cite{ShenULBD03,RalphG91,KeilD96}. This has
important practical implications for THz emitter design.

Such ``anode-enhanced'' THz emission has been explained previously
by a trap-enhancement of the electric field near the anode
\cite{RalphG91} or by the presence of a high field region near
the anode in the case of a Schottky contact \cite{KeilDKD94}. However,
experimental measurements \cite{KeilD96} suggest that the asymmetry
between two Ohmic electrodes may be attributed to the difference of
mobilities between electrons and holes.

As discussed above, the simulation treats the
case of PCS electrodes that are each screened to a similar extent before
photoexcitation. In reality, depending on the substrate and type of
contacts, some asymmetry in this screening may exist. However here we show that the
presence of contact asymmetry is not a necessary condition in order to
observe ``anode-enhanced'' THz emission.

\begin{figure}
\begin{center}
\includegraphics[width=8.6cm]{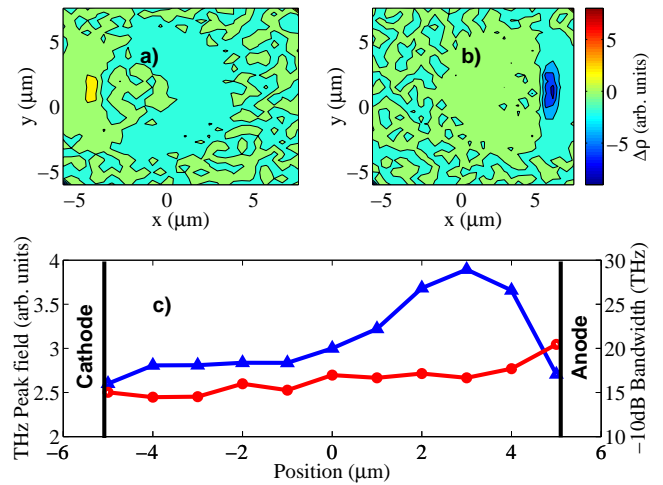}
\end{center}
\caption{ (Colour online) Simulation results showing ``anode
enhanced'' THz emission. (a) and (b) are contour plots of the change
in charge density, $\Delta\rho$, for simulations where the laser
pulse arrived near the cathode ($x=-3\,\rm{\mu m}$) and anode
($x=3\,\rm{\mu m}$) respectively.  $\Delta\rho$ is the change in
charge density between $t=0$ to $t=300\,\rm{fs}$ projected onto the
$x$--$y$ plane. $\Delta\rho$ was averaged over all slices in the
$z$-direction. A summary of 11 simulations is shown in (c) where the
peak THz electric field (triangles) and FWHM (circles) are plotted
as a function of laser spot position.  } \label{LaserPosition}
\end{figure}

This effect is most easily illustrated by plotting the movement of
charge over the duration of the simulation.  The contour plots in
Fig.~\ref{LaserPosition}(a) and (b) show the change in charge density,
$\Delta\rho=\rho(300\,\rm{fs})-\rho(0\,\rm{fs})$, as a function of
$x$--$y$ position of the $10\, {\rm \mu m}$ gap PCS.  Plots (a) and
(b) are extracted from a simulation where the PCS was illuminated near
the cathode and anode respectively. It can be seen that the displaced
charge when the PCS is illuminated near the anode is considerable
larger.

The enhancement effect can be understood in terms of the higher
effective mobility of the photo-injected electron population over the
hole population. When photoexcitation occurs close to the anode
electrons (holes) migrate into the high (low) field region (see
Fig.~\ref{Screening}).  Therefore, in this case it is the more mobile
charges that are affected by the higher electric fields. Thus when
electron-hole pairs are generated near the anode the higher electron
mobility results in an increase in the effective current density, and
as a consequence a larger amplitude THz electric field is emitted.

The above explanation is somewhat simplified, as under typical electric
fields $\rm{\Gamma} \rightarrow \rm{L}$ inter-valley scattering will lead to a drop in electron
mobility. However, as the simulation considers intervalley scattering, all
results presented here include this additional effect.

A set of simulations were performed where the laser pulse arrival
position was varied across the $10\, {\rm \mu m}$ gap of the PCS. A
summary of these data is shown in Fig.~\ref{LaserPosition}(c), where
the peak electric field is plotted as a function of the distance of
the laser pulse from the center of the PCS gap.  The peak THz field
is enhanced by approximately 40\% as the pulse approaches the edge
of the anode.  The field drops again nearer the anode as a larger
fraction of the laser beam-waist is incident outside the active region
of the PCS. These results are consistent with previous experimental
observations.\cite{ShenULBD03,RalphG91,KeilD96}

The widths of the Fourier transforms at -10\,dB of the signal at
different positions are also plotted in Fig.
\ref{LaserPosition}(c). It is interesting to note that the bandwidth
increases as the excitation nears the anode. This can be attributed to
the fact that in that region the electric field is greater, resulting
in a shorter THz transient, as discussed in Section
\ref{s:bias}.

Finally, it is useful to characterise the timescale of space-charge
screening during the photo-excitation of a PCS, so that the influence
of pump-pulse duration on the THz emission process can be analysed.
We fitted an error function to the $x=\pm 2.5\,\rm{\mu m}$ position
slices of the graph shown in Fig.~\ref{Screening} and chose to define
the screening time as 2 standard deviations of that error
function. The screening time for this simulation was found to be $\sim
100\,\mathrm{fs}$ for all simulated pump pulses between 10 and
100\,fs. This suggests that space charge screening is dominated by the intrinsic dynamics of
carriers in the semiconductor and it is not significantly affected by
the carrier generation time.

\section{Conclusions} \label{s:conclusion}

A semi-classical Monte Carlo model that simulates the carrier
dynamics in a photoconductive switch was presented. The THz
transient has been shown to narrow with the applied voltage, as
well as to be strongly dependent on the pulse width of the
exciting laser. The emitted THz power was also studied, and was
found to have a maximum for pump pulses of $40\,\rm{fs}$ duration.
A drop in power for shorter pulses was explained by a broadening
in the energy distribution of the exciting pulse, which starts to
contain a significant number of photons with sub-band-gap and over
L-valley gap energies. This reduces the quantity of free
photo-carriers and increases their average effective mass. The
reduction in THz power for longer excitation pulses is explained
by a decrease in the bias field as a result of screening, since
the photo-carriers are injected over a longer period of time.

The screening of the electric bias field was studied. It was found
that the screening time is independent of the pump pulses (for
pulses under $100\,\mathrm{fs}$), implying that the screening time
is mainly dependent on the carrier dynamics and not on their
injection time on this timescale. The electric field was found to
be screened over the gap between the electrodes of the PCS, but
not near the edges of the electrodes. This fact combined with the
difference of mobilities of electrons and holes produces an
increase in the peak THz electric field and a broadening in
bandwidth when carriers are photo-excited near the anode edge.

In the future this model promises to be useful for testing new PCS
electrode designs and substrate materials (for example low temperature
grown GaAs or ion-implanted semiconductors).  For better correlation
with experimental results it would also be informative to couple a
full three-dimensional electromagnetic simulation with the existing
carrier dynamics simulation.

\section{Acknowledgements}
The authors would like to thank the EPSRC (UK) and the Royal Society
for financial support of this work, ECC wishes to thank CONACyT
(M\'exico) for a scholarship.

\bibliographystyle{apsrev}

\end{document}